\documentclass{tlp}

\usepackage{latexsym}
\usepackage{amssymb}
\usepackage{amsmath}
\usepackage{bsymb}

\usepackage{graphicx}
\usepackage{paralist}

\newcommand{\prob}{{\sc ProB}}

\newcommand{\ignore}[1]{}

\newtheorem{example}{Example}[section]

\begin{document}
\title{Constraint-Based Deadlock Checking of High-Level Specifications}

\author{Stefan Hallerstede, Michael Leuschel\\
 Institut f\"{u}r Informatik, Universit\"{a}t D\"{u}sseldorf\\
  Universit\"{a}tsstr. 1, D-40225 D\"{u}sseldorf\\
  E-mail: \{ halstefa, leuschel \}{@cs.uni-duesseldorf.de}
}

\maketitle

\begin{abstract}
Establishing the absence of deadlocks is important in many applications of formal methods.
The use of model checking for finding deadlocks in formal models is 
often limited.
%limited because in many industrial applications the state space is either infinite or much too large to be explored exhaustively.
%An alternative approach is to try and prove the absence of deadlocks.
%The resulting proof obligations are, however, very big and difficult to prove.
In this paper we propose a constraint-based approach to finding deadlocks employing 
the \prob\ constraint solver.
% to find values for the constants and variables of formal models that describe a deadlocking state.
We present the general technique, as well as various improvements that had to be performed
 on \prob's Prolog kernel, such as reification of membership and arithmetic constraints. % boolean constraint solver with reification
This work was guided by an industrial case study, where a team from Bosch was modeling a cruise control system.
Within this case study,
 \prob\ was able to quickly find counter examples to very large deadlock-freedom constraints.
% a formula of about 900 partly nested conjunctions and disjunction among them 80 arithmetic 
% and 150 set-theoretic predicates (in total a formula of 30 pages), within a few seconds.
In the paper, we also present other successful applications of this new technique.
Experiments using SAT and SMT solvers on these constraints were thus far unsuccessful.
 %and too large to load into Rodin tool
\end{abstract}
% --------------------------

\section{Introduction}

Formal modelling of discrete event systems is an important tool in order to verify their correct functioning.
Among others we may want to verify 
\begin{inparaenum}[(a)]
\item termination of certain components, \label{it:termprop}
\item avoidance of unsafe states or \label{it:safeprop}
\item absence of deadlocks. \label{it:deadprop}
\end{inparaenum}
In the formal models discussed in this article property (\ref{it:deadprop}), absence of deadlocks, is considered the principal property.
In fact, in the industrial applications that have motivated the work described in this paper 
the first two properties (\ref{it:termprop}) and (\ref{it:safeprop}) play only a small role.
 
The classic approach to locating deadlocks is model-checking.
Model-checking can provide fast feedback, but is also associated with known problems:
in many applications the state space is either infinite or much too large to explore exhaustively.
Furthermore, model checking is particularly problematic when the out-degree of certain states is very large.

In this paper we describe a successful application of constraint solving to verify absence of deadlocks.
In the industrial case study that has started this work, a team from Bosch attempts to develop a deadlock-free formal model of a cruise control system.
For this application, constraint solving typically finds counter examples to 
deadlock-freedom constraints of more than 30 A4 pages in under two seconds, while model checking was unsuccessful.
It turns out that this approach ---on top of being much less sensitive to the size of the state space--- 
has additional benefits.
It exploits safety properties that have been specified (having the positive side effect of encouraging their specification)
and it can be easily related to verification by formal proof.
Model-checking can succeed showing absence of deadlocks even if this cannot be verified by formal proof using all specified safety properties.
Constraint solving will only succeed if also a proof can be found: it is not based on model execution.
In Event-B \cite{Abrial10}, the formal method we have used in the case studies, this helped to achieve a more comprehensive methodology of verification.
In the end, it is the mix of constraint solving, model checking and proof that advanced the case study of Bosch using \prob{} \cite{DBLP:journals/sttt/LeuschelB08} and Rodin \cite{journals/sttt/AbrialBHHMV10}.
Rodin is a software tool for formal modelling with Event-B\@. %\marginpar{last sentences added}
It is equipped with editors, a proof obligation generator and some theorem provers. 
\prob{} is also available as a plug-in for Rodin providing animation, model-checking and constraint-solving facilities.

\subsection{Deadlock-Freedom in Event-B}

We discuss deadlock freedom in terms of Event-B\@.
However, the results are not specific to Event-B\@.
The concept of deadlock freedom applies quite universally to state-based formal methods.
In fact, the corresponding constraint solving technique implemented in \prob{} can be 
applied immediately to models created using the ``classical'' B-Method \cite{Abrial96}
 and the Z specification notation. % \cite{Spivey:Z}.

We only present the concepts of Event-B necessary to discuss deadlock freedom.
In particular, we ignore concepts such as refinement, theorems or witnesses.
% as they would distract from the core contribution of the paper.
An Event-B model is called a {\em machine}.
A simple machine is shown in Figure~\ref{fig:example1}.
\newcommand{\mmin}{\mathit{min}}
\begin{figure}[htp]
\begin{footnotesize}
\begin{center}
$
\begin{array}{@{}l@{}}
  \mathrm{MACHINE}\ \mathit{MinSet}\\
  \mathrm{CONSTANTS}\ N\\
  \mathrm{AXIOMS}\ N \subseteq 0 \upto 3 \land N\neq \emptyset\\
  \mathrm{VARIABLES}\ s, \mmin, z\\
  \mathrm{INVARIANTS}\ s \subseteq 0 \upto 3 \land \mmin \in 0 \upto 3 \land z \in 0 \upto 4\\
  \mathrm{EVENTS}\\ \quad
  \begin{array}{@{}l@{}}
    \mathrm{INITIALISATION}\ =\ s \bcmeq N\cup\{3\} \parallel \mmin \bcmeq 3 \parallel z \bcmeq 4\\
    \mathit{acc}\ =\ 
    \mathrm{ANY}\ x\ 
    \mathrm{WHEN}\ \mmin \in s \land x \in s \land x < \mmin \ %\\ 
    %\phantom{\mathit{acc}\ =\ \mathrm{ANY}\ x}\ 
    \mathrm{THEN}\  s \bcmeq s-\{\mmin\}   \parallel \mmin \bcmeq x\ 
    \mathrm{END}\\
    \mathit{rej}\ =\ 
    \mathrm{ANY}\ x\ 
    \mathrm{WHEN}\ \mmin \in s \land x \in s \land x > \mmin\ 
    \mathrm{THEN}\ s \bcmeq s-\{x\}\
    \mathrm{END}\\
    \mathit{get}\ =\ 
    \mathrm{WHEN}\ s = \emptyset\ \mathrm{THEN}\ z \bcmeq \mmin\ \mathrm{END}
  \end{array}\\
  \mathrm{END}
\end{array}
$
\end{center}
\end{footnotesize}
\caption{\label{fig:example1}A machine for computing the minimum $z$ of a set $s$}
\end{figure}
The state of a machine is described in terms of {\em constants\/} and {\em variables}.
The possible values of the constants are constrained by {\em axioms\/} $A\,=\, A_1\land\ldots\land A_r$%
\footnote{All indices in this paragraph have the range ``$\geq 0$''.} and the possible values of the variables
by {\em invariants} $I\, =\, I_1\land\ldots\land I_s$, all expressed in first-order predicate logic augmented with arithmetic over integers and (typed) set theory.
State changes are modelled by {\em events}.
Each event consists of a collection of 
{\em parameters} $p_1,\ldots,p_i$, of
{\em guards} $g \,=\, g_1 \land\ldots\land g_j$ and of 
{\em actions} $a$ (usually a collection of simultaneous update statements $a_1\parallel\ldots\parallel a_k$).%
\footnote{The exact form of the update statements $a_\ell$ is not relevant for this article.} 
Guards are predicates over the constants, variables and parameters.
We use the following schema to describe events:
%\begin{align*}
$
  \mathrm{ANY}\ p_1,\ldots,p_i\ \mathrm{WHEN}\ g\ \mathrm{THEN}\ a\ \mathrm{END}$.
%\end{align*}
We leave out clauses of an event that are ``empty''.
For instance, an event without parameters is written $\mathrm{WHEN}\ g\ \mathrm{THEN}\ a\ \mathrm{END}$;
and an event without parameters and guards consists just of the actions $a$.
An event needs to be enabled to change the state as described by its actions.
An event is {\em enabled\/} in a state if there are values $p_1,\ldots,p_i$ that make its guard $g$ true in that state.
We denote the {\em enabling predicate} $(\exists p_1,\ldots,p_i\qdot g)$ of an event $e$ by $G_ e$.
Being enabled an event {\em can\/} be executed by performing all its actions simultaneously.
A special event, called the $\mathrm{INITIALISATION}$ is executed (once) first to initialise the machine.
The $\mathrm{INITIALISATION}$ event does not have guards or parameters.

\subsection{Constraint Solving of Deadlock-Freedom Proof Obligations}
\label{sec:csolvdlk}

A state of a machine in which none of the events (except for the $\mathrm{INITIALISATION}$ event) is enabled is called a deadlock.
We can search for such states by \emph{model checking}, namely by looking at all the reachable states and enabled events.
Another approach is to \emph{prove} absence of deadlocks.
The invariant of a machine describes a superset of the reachable states.%
\footnote{This in turn can be verified by model checking or proof.}
So, if the invariant is ``precise'' enough it should imply that always one of the events is enabled.
Formally this can be expressed in terms of the \emph{proof obligation}:
\begin{align}
 A\land I \,\limp\, G_{e_1} \lor\ldots\lor G_{e_n} \tag{DLF}
\end{align}
where $A$ are the axioms, $I$ are invariants and $G_{e_\ell}$ ($\ell\in 1\upto n$) the enabling predicates of the events $e_\ell$.
Counterexamples to the proof obligation can be found using \emph{constraint solving}.

Now we have three approaches to finding out about deadlocks: model checking, proof and constraint solving.
Often they do not yield the same results.
Model checking finds only those deadlocks that can actually occur during execution of the events.
Proof and constraint solving signal deadlocks depending on whether the proof obligation holds.
If attempting to prove it, we may ``get stuck'' in a proof.
This may happen because the proof obligation cannot be proved (i.e.\ the invariant is too weak or the enabling predicates are too strong)
or because something is wrong with the proof.
These two causes are difficult to distinguish for complicated proof obligations like the afore-mentioned 30 A4 pages.
Constraint solving produces a counter example if the implication does not hold.
Hence, it helps distinguishing the two causes.
Although proof applies, in general, to a much larger class of formulas (that is, proof obligations) than constraint solving
we found that many models we encountered use only a restricted class of formulas where constraint solving could be applied, too.

Model checking the Event-B machine of Figure~\ref{fig:example1} detects a deadlock for the state
$N=0\upto 3 \land s=\{0\} \land \mmin =0 \land z=4$: 
if the set $s$ contains only one element, none of the events is enabled.
We change the guard of event $\mathit{get}$ to $s=\{\mmin\}$ to correct the problem.
Now model checking succeeds -- there is no deadlock.
However, we cannot prove this. Why? 
The deadlock-freedom proof obligation is the following:
\begin{align*}
  \begin{array}{@{}l@{}l@{}}
  & N \subseteq 0 \upto 3 \land N\neq \emptyset \land s \subseteq 0 \upto 3 \land \mmin \in 0 \upto 3 \land z \in 0 \upto 4\\
  \limp\ & (\exists x\,\qdot\,\mmin \in s \land x \in s \land x < \mmin)\, \lor \\ & (\exists x\,\qdot\,\mmin \in s \land x \in s \land x > \mmin) \lor s=\{\mmin\}
  \end{array}
\end{align*}
Constraint checking of the corrected machine yields a deadlock in the state
$N=\{3\} \land s=\emptyset \land \mmin =0 \land z=0$:
neither $\mmin \in s$ nor $s=\{\mmin\}$ holds in this state.
Adding $\mmin \in s$ to the invariants of the machine solves the problem.
We have discovered a fact about our model ---the minimum to be computed is always contained in the set $s$---
and we have specified this fact as an invariant describing the reachable states.
Doing this kind of analysis exclusively by means of proof on large proof obligations can be very difficult.
Model checking and constraint solving make it practically feasible to analyse such proof obligations.

Due to the number of constants and variables in realistic models, model checking also encounters a number of known problems
that can be avoided using constraint solving.
For instance, there can be a practically infinite number of ways to instantiate the constants of a B model.
In this case, model checking will only find deadlocks for the given constants chosen. 
And the number of choices is exponential in the number of constants.
Constraint checking proceeds smarter, for instance, by propagating constant values 
(but remains worst-case exponential, of course).

%\marginpar{(Similarly, if the invariant can be violated then the CBC model checker may fail to find
% a deadlock which the model checker can find (by visiting states which violate the invariant)).}
 
\subsection{Constraint Solving with \prob}
\label{sec:probintro}

\prob\ \cite{DBLP:journals/sttt/LeuschelB08} is a validation tool for high-level specification formalisms, 
such as the B-Method, Event-B, Z and CSP\@.
\prob\ provides various validation techniques, such as animation, model checking, constraint checking, refinement checking and test-case generation.
The various specification formalisms are encoded in Prolog in the form of an interpreter, usually encoding an operational semantics of the language.
For example, CSP is embedded within \prob\ in the form of Roscoe's operational semantics \cite{Roscoe:book}.
The foundation of the B-Method, Event-B and Z are set theory, (integer) arithmetic and predicate logic.
As such, \prob\ provides constraint solving over sets and derived datatypes such as relations and functions.

The basic constraint solving functionality concerns 
\begin{inparaenum}[(a)]
\item checking for invariant preservation by all or by some specific operations, \label{it:prob:ip}
\item validating data only available at deployment time with respect to formal properties used during development \label{it:prob:dv}
\item finding some state satisfying given axioms and invariants and, finally, \label{it:prob:fs}
\item finding a deadlock. \label{it:prob:dlk}
\end{inparaenum}
The first (\ref{it:prob:ip}) is similar in functionality to Alloy \cite{Jackson:Alloy} and has already been discussed in \cite{DBLP:journals/sttt/LeuschelB08}.
The second (\ref{it:prob:dv}) has been successfully applied by Siemens to analyse railway networks in production \cite{LeuschelEtAl:STS09}.
The third (\ref{it:prob:fs}) is useful to check axioms and invariants for contradictions.
The last (\ref{it:prob:dlk}) is described in more detail in this article.
In addition to the deadlock checking discussed in Section~\ref{sec:csolvdlk},
a variant is supported by \prob\ that permits specifying a predicate of interest $P$.
Using this we can restrict deadlock checking to a subset of states that may provide further insight.
For instance, in the example discussed above we could have taken $P$ to be 
$\mmin \in s$ first, in order to see whether it is sufficient to achieve deadlock-freedom.

% --------------------------

\section{Principles of Constraint-Based Deadlock Checking}

In this section we sketch the algorithm implemented for constraint-based deadlock checking in \prob.
First we discuss the direct approach that addresses directly proof obligation (DLF) by negating the guards (DLN).
Based on some criticism of the direct approach we finally present the promised algorithm in Section~\ref{sec:improved}.

\subsection{Direct Approach}

The direct approach is quite simple: construct a formula (DLN) consisting of the conjunction of
the axioms $A$ of the model,
the invariants $I$ of the model and
the negation of the enabling predicate ($\neg G_{e_\ell}$) for every event of the model.
Formally,
\begin{align}
 A\land I \land \neg G_{e_1} \land\ldots\land \neg G_{e_n} \tag{DLN}
\end{align}
If we find a solution for this formula, then we have found a deadlocking state.
As discussed in Section~\ref{sec:csolvdlk} this state is not necessarily reachable from the initial states.
However, this state is allowed by the axioms and invariants of the model. 
Any attempt at proving the deadlock-freedom proof obligation (DLF) is guaranteed to fail.
(The model should thus be corrected, independently of whether this state can actually be reached or not.)
Note that when the axioms of the model are inconsistent, the constraint solver will not be able
to find a valid valuation of the constants and thus (by monotonicity) also not find a deadlocking state.
Consistency of the axioms can be checked by \prob, 
using the technique (\ref{it:prob:fs}) of Section~\ref{sec:probintro} or %``find some state satisfying\ldots'' or 
simply by starting an animation of the model.

%\begin{figure}
%\begin{center}
%  \includegraphics[width=7.5cm]{Figures/DeadlockCBCAnalysis} %original: {width=175pt, angle=270}
%  \includegraphics[width=7.5cm]{Figures/DeadlockMCAnalysis} %original: {width=175pt, angle=270}
%  \caption{Finding Deadlocks using Constraint Solving and using Model Checking}\label{Overview}
%\end{center}
%\end{figure}

% --------------------------

{\bf Criticism of the Direct Approach}
\label{sec:critic}
Applying the \prob\  constraint solver to the constraint (DLN) above yields
 a counter-example to the deadlock-freedom proof obligation (DLF) if the constraint (DLN) is satisfiable.
However, the above approach suffers from a series of shortcomings that restrict its potential use:

%\begin{description}
% \item[Redundancy.]
\textbf{Redundancy.}\quad
  \prob\ will find values for {\em all\/} variables and constants of the model. 
 However, it is quite common that some of the variables and constants are not relevant for the guards of the
 events (e.g., they are only used in the action parts or are sometimes only there for helping with the proof effort
  but do not affect the behaviour of the model).
For example, in the machine of Figure~\ref{fig:example1} variable $z$ is not relevant for the guards.

\noindent\textbf{\itshape Solution.}
 To solve this we partition the formula  
  into connected sub-components. We can ignore any sub-component not related to any guard.
  
% \item[Inaccuracy.] 
 \textbf{Inaccuracy.}\quad
 Sometimes the user is not interested in arbitrary deadlocks, but only in
  a certain class of deadlocks.
  For example, in when analysing the machine of Figure~\ref{fig:example1} 
  we may only be interested in looking for deadlocks in states where $\mmin\in s$.
 
\noindent\textbf{\itshape Solution.}
  To address this requirement,
     we give the user the possibility to specify an additional predicate $P$ of interest.
   This predicate is added to the constraints to be solved.
   In addition, we optionally filter out any event that can obviously not fire given $P$;
   for example, any event that has {\tt Counter=5} in its guard given $P=$ {\tt Counter=10}.
   This obviously simplifies the constraint to be solved, and more importantly, can sometimes result in a much
   better decomposition of the formula into independent sub-components.
 
% \item[Inefficiency.] 
 \textbf{Inefficiency.}\quad
 Formulas may not directly fit shapes that can be treated efficiently by the constraint solver.
 For instance, the use of the existential quantifiers in enabling predicates complicate the constraint solving process.
 Indeed, the \prob\ kernel will usually wait until all quantities used inside the existential quantifier are known
  before evaluating it; see Section~\ref{constraint-solving}.
 % However, often existential quantifiers only refer to a subset of the guards or can be completely removed.
  
\noindent\textbf{\itshape Solution.}
  To solve this we run a simplifier on the enabling predicates before adding them to the constraint store.
  For example, the predicate $\exists x\,\qdot\, \mmin\in s \land x\in s$ 
  can be simplified to $\mmin\in s \land (\exists x\,\qdot\, x\in s)$ and further to $\mmin\in s \land s\neq \emptyset$.
  Comparing this to the machine of Figure~\ref{fig:example1} one can see that simplification will often not produce rewritings to this degree.
  However, in the case studies described later it was very effective.
  Currently, the simplification process is straightforward, mainly
  addressing common patterns that appear in guards, such as:
\begin{itemize}
 \item $\exists x. x\in S$  is simplified to $S\neq \emptyset$;
 \item $S \neq \emptyset$  is simplified to $\True$,  in case $S$ is guaranteed to be non-empty;
  \item $\exists x. x>E$  is simplified to $\True$;
 \item $\exists x.(x=E \wedge P)$ is simplified to $P[E/x]$.
\end{itemize}
A couple of simplifications have been implemented in Prolog.
Outside of \prob{} we have also experimented with good results with the simplifier of the theorem prover Isabelle \cite{Paulson94} 
that has a large number of simplifications that come with its theories of set theory and arithmetic.
Because in our methodology theorem proving already plays an essential role, 
we could reuse a simplification already implemented. 
Integration with theorem prover looks like a promising option for the future.
%
%\end{description} 

\subsection{Improved (More Efficient) Algorithm}
\label{sec:improved}

The discussion of Section~\ref{sec:critic} suggests the improved algorithm for deadlock checking shown in Fig.~\ref{fig:dlkchkalg},
the algorithm implemented in the current version of ProB (1.3.3):
\begin{figure}[ht]
\begin{enumerate}[\quad\ 1:]
 \item {\bf input}: predicate $P$ of interest, list $L$ of events of interest 
 \item AI := $A$ $\wedge$ $I$ $\wedge$ $P$; ~~~ (* axioms, invariants and predicate of interest *)
 \item Deadlock := $\True$; ~~~ (* only relevant enabling predicates are considered *)
 \item {\bf for} each event $e$ in L {\bf do}:
 \item ~~~ extract enabling predicate $G_e$ of event $e$;
 \item ~~~ simplify $G_e$; ~~~ (* as described above *)\label{prog:simp}
 \item ~~~ {\bf if} solve(AI $\wedge$ $G_e$) $\neq$ false {\bf then} 
  \item ~~~ ~~~ (* otherwise the event is always disabled given P *)
 \item ~~~ ~~~ Deadlock := Deadlock $\wedge$ $\neg(G_e)$
 \item ~~~ {\bf fi}
 \item {\bf od}
 \item sort conjuncts nested inside Deadlock (move most-used conjuncts to the front)
 \item $\langle C_1,\ldots,C_n\rangle$ := components(AI $\wedge$ Deadlock); \label{prog:part} %\footnotemark
 \item {\bf return} $\langle$solve($C_1$),\ldots,solve($C_n$)$\rangle$ 
\end{enumerate}
\caption{Improved algorithm for deadlock checking\label{fig:dlkchkalg}}
\end{figure}
%The algorithm improves on the direct approach presented earlier by addressing the problems stated in Section~\ref{sec:critic}.
The problem of redundancy is addressed by the partitioning in line~\ref{prog:part};
the problem of inaccuracy by the incorporation of the predicate of interest;
and the problem of inefficiency by the invocation of the simplifier in line~\ref{prog:simp}.
Note that the latter two techniques are orthogonal to the algorithm and could be applied to other constraint checking problems in the same way.
Finally, in line~\ref{prog:part} we can optionally remove all components $C_i$ not relevant for the deadlock.
% FOR OTHER ARTICLE:
%extraction of guards: parameter specifies whether we want to extract predicates
% from becomes such that or becomes element of.
% (Default: no, as these are dealt with by FIS proof obligation.)
% \marginpar{One could check this separately by constraint checking?}
 
%\footnotetext{Optionally: remove all components $C_i$ not relevant for Deadlock.} % get footnote of figure fig:dlkchkalg on right page...

% 
%
%
%Explain why existential quantifiers decrease performance of constraint solving
%Note: we can chose to ignore wd-issues, deciding that here we try to find deadlocks, not WD-problems.%
%\marginpar{Chose for the matter of this article? Or can the user chose? I'd prefer not to discuss wd at all.}
%Any counter-example found is guaranteed to be a well-defined deadlock.
% 
%explain basics of counting and sorting
%
%explain basics of component partitioning; explain different rule for existential quantifiers compared to axioms

% --------------------------

\section{Core of the \prob\ Constraint Solver}
\label{constraint-solving}

In this section we present some key aspects of the \prob\ constraint solver and its implementation in Prolog.
We pay particular attention to the extensions that were made in light of the Bosch application (Section~\ref{bosch}).

\subsection{Overview}

The \prob\ kernel provides a constraint-solver for the basic datatypes of B (and Z) and the various operations on it.
As such, it supports booleans, integers, user-defined base types, pairs, records and inductively: sets,
 relations, functions and sequences.
These datatypes and operations are embedded inside B (and Z) predicates, which can make use of the
 usual connectives ($\wedge, \vee, \Rightarrow, \Leftrightarrow, \neg$) and typed universal ($\forall x. P \Rightarrow Q$)
  and existential ($\exists x . P \wedge Q$) quantification.

%An overview of the \prob\ kernel is shown in Figure~\ref{ProB_CBC_Kernel}.
%As can be seen,
 \prob\ integrates various constraint solvers in its kernel:
\begin{compactenum}[\hbox{$\bullet$}]
\item
Integers are represented using Prolog integers.
To implement arithmetic constraints, \prob\ uses the SICStus CLP(FD) finite-domain library \cite{CarlssonOttosson:PLILP97}.%
%\footnote{For efficiency reasons, \prob\ tries to use CLP(FD) only where necessary. There is also always
% a pure Prolog backup version available; see below.}
 
\item Elements of basic sets are represented internally
  as terms of the form {\tt fd(Nr,T)}, where T is the type name and Nr is the number of the element.
 Equality is then implemented simply using unification. Disequality is generally
  implemented using the disequality operator of CLP(FD).
Thus, disequality can also sometimes deterministically instantiate its arguments.
E.g., given the user-defined type {\tt S=\{a,b\}}, the predicate {\tt x /= a} will force the value of {\tt x} to be {\tt b}.

\item The constraints for the more complicated types have been written in Prolog with co-routines.
Note that \prob\ employs various set-representations: AVL-trees for fully known sets (to be able to deal with
 large sets arising in industrial applications, cf. \cite{LeuschelEtAl:STS09}), closures to represent certain sets symbolically
  and Prolog lists for partially known sets.
A feature that distinguishes \prob\ is that it not only deals with simple sets, but also allows sets of sets, relations, etc.

Generally, co-routines are used to block non-deterministic computations. A non-deterministic computation
  provides an estimate of the number of solutions to the \prob\ waitflags store and obtains
 a Prolog variable on which it can block (called a waitflag).
This waitflag will be instantiated by the enumeration process, which will unblock computations
 with the least number of solutions first and will also take care of labeling the
 CLP(FD) variables.
\item Finally, the boolean predicate solver is also written in Prolog using co-routines.
We describe some aspects of its implementation in more detail below.
We did not reuse the SICStus boolean constraint solver mainly due to the treatment of
 undefined predicates and in order to link the labeling with our other solvers.
\end{compactenum}

\ignore{***
\begin{figure}
\begin{center}
  \includegraphics[width=7.5cm]{Figures/ProB_CBC_Kernel} 
  \caption{A view of the \prob\ kernel}\label{ProB_CBC_Kernel}
\end{center}
\end{figure}
  ***}
 
\subsection{Challenges in Reusing CLP(FD)}

Initially \prob\ did not use CLP(FD): up until version 4.0.8,
 SICS advised against combining co-routines and
 CLP(FD) in the same program.
With the arrival of SICStus 4.1, we started to integrate CLP(FD) into the \prob\ kernel.
Still, we encountered segmentation faults in the first versions (SICStus 4.1.1).
These issues have been fixed in 4.1.2.

In B and Z, integers are unbounded but B also provides the {\em implementable} integers, which in typical
industrial applications fall in the range $-2^{32} \upto 2^{32}-1$.
Unfortunately,
 the CLP(FD) library by SICStus can only represent integers from $2^{28}\upto 2^{28}-1$ in 32-bit mode.
Hence, \prob\ always has a Prolog ``backup'' solution (without interval propagation) available
 and tries to catch overflows when posting CLP(FD) constraints.
Still, overflows can happen outside of the control of the kernel, simply by instantiating a variable.
Hence, \prob\ can also be run completely without CLP(FD).
Another solution is using the 64-bit version of SICStus Prolog, where we can
 handle integers in the range $-2^{60} \upto 2^{60}-1$.%
 \ignore{**
\footnote{This is only a solution as long as the industrial applications themselves do not require
 64-bit integers in the models.}
**}

\ignore{***
\begin{footnotesize}
\begin{verbatim}
| ?- X #= 1024*1024*256.
! Representation error in argument 2 of user: #= /2
! CLPFD integer overflow
! goal:  _102#=268435456
 | ?- X #= 1024*1024*256-1.
X = 268435455 ? 
| ?- X #= -1024*1024*256.
X = -268435456 ? 
yes
| ?- X #= -1024*1024*256-1.
! Representation error in argument 2 of user: #= /2
! CLPFD integer overflow
! goal:  _102#= -268435457
\end{verbatim}
\end{footnotesize}
**}

Difficulties arise when the models contain unbounded mathematical integers.
In the Bosch application, we have several mathematical integers (e.g., to represent time)
 and the constraint solver may be asked to solve a constraint
 {\tt x>y} and {\tt y>=x}.
This situation actually arises very frequently in constraint-based deadlock checking,
 when events have common guards or complementary guards.
Unfortunately, CLP(FD) does not deal very well with such constraints.
First, it does not detect an inconsistency after posting {\tt\verb+X#>Y,Y#>=X+}.
Second, if we later add another constraint, such as
 {\tt\verb+Y#>200+} we get an integer overflow error.%
 \footnote{Using the hardware configuration of Section~\ref{sec:casestudy} this happens after about 40 seconds.}
\ignore{***
\begin{footnotesize}
\begin{verbatim}
| ?- X #>Y, Y#>=X, Y#>200.
! Representation error in user:'t=<u+c'/3
! CLPFD integer overflow
! goal:  't=<u+c'(_107,_108,-1)
\end{verbatim}
\end{footnotesize}
**}
Our solution is to add a time-out when posting constraints, and revert to the Prolog
backup if a time-out occurs.
Furthermore, we have extended our boolean constraint solver to detect identical atomic predicates.
Basically, every atomic predicate is normalised and then checked if it occurs at another place
 in the same formula: if it does, the predicate is evaluated only once.
As a special case, it detects the inconsistency above.
% Michael : can we infer that (x>y & y>=x) is always false !
 % and the reification\marginpar{reification explained below} variable reused.

Finally, CLP(FD) does not deal with undefinedness \cite{DBLP:conf/cp/FrischS09} the same way that B does :
 {\tt\verb+X in 1..10, X/0#=10+} simply fails, while in B this is an erroneous formula.
\prob\ tries to catch those errors. This is reflected inside the boolean constraint solver  and the fact
  that we do not yet use CLP(FD) for division and modulo.

%well-definedness problems: catch and raise errors,... cite Stuckey paper 

 \ignore{***
 
\subsection{Implementation}

 The boolean constraint solver:
 \begin{itemize}
  \item we can ask the kernel to make a given predicate $P$ true
  \item we can ask the kernel to ensure that a given predicate $P$ is false
  \item we can reify $P$: represent the truth status of $P$ by a variable:
   if the \prob\ kernel can determine whether $P$ is definitely true or false then
    the variable is instantiated accordingly; similarly, if the variable
    becomes instantiated, the predicate is forced to be either true or false (reverting to items
    1 and 2 above)
  \end{itemize}
 
 In mode three: reification of basic constraints has to be available; currently
  equality, disequality, arithmetic comparisons, membership, non-membership.
 $\subseteq$ not yet reified. universal, existential quantification not yet reified, but sometimes expanded
 into conjunction respectively disjunction.
 ****}
 
\subsection{The \prob\ Boolean Constraint Solver}

%\subsubsection{Predicate Truth Values}

The boolean constraint solver uses {\em reifications\/} of the basic atomic predicates
 to communicate with the other solvers. % of Figure~\ref{ProB_CBC_Kernel}.
More precisely, given a basic atomic predicate $P$, we associate with it a Prolog variable $R_P$.
If another solver can determine that $P$ must be true, then it sets $R_P$ to {\tt pred\_true}.
If it can determine that $P$ must be false, then it sets $R_P$ to {\tt pred\_false}.
Similarly, if $R_P$ is set to {\tt pred\_true} (resp.  {\tt pred\_false})
 by the boolean constraint solver, the other solver
 should add $P$ (resp. $\neg P$) to its constraint store.

The boolean constraint solver also uses reification internally to treat more complex subformulas.
Figure~\ref{equivalence} shows, e.g., how the equivalence connective is implemented
inside the {\tt b\_check\_boolean\_expression2} predicate.
The variable {\tt LR} is the reification of the left-hand side predicate {\tt LHS}.
Similarly, {\tt RR} is the reification of the right-hand side predicate {\tt RHS}.
Finally, {\tt Res} is the reification of the equivalence {\tt LHS <=> RHS}.
The {\tt equiv} procedure will ensure consistency of the three reification states
%\marginpar{{\tt equiv} is the clause for ``{\tt <=>}'' ?}
 and ensure propagation of information: it blocks until at least one of the 
 reification variables is instantiated and then propagates the information,
 possibly using the auxiliary {\tt negate} procedure.
For example, if {\tt Res} and {\tt LR} are known to be false ({\tt pred\_false}),
 then {\tt LL} will be forced to {\tt pred\_true}.%
\footnote{In \prob\ 1.3.4 we have introduced a more refined implementation of {\tt negate}
 using attributed variables. It will, e.g., infer from {\tt negate(X,Y),negate(Y,Z)}
  that {\tt X==Z}.}
This will trigger further information propagation inside the call
   {\tt b\_check\_boolean\_expression} for {\tt LHS}.
E.g., if {\tt LHS} is {\tt x=2} then the Prolog variable representing
 the B identifier x would be forced to 2 (actually {\tt int(2)}).
The code for the other connectives in {\tt b\_check\_boolean\_expression2} 
 is a bit more complicated, due to the treatment
 of undefinedness.

\begin{figure}
\begin{footnotesize}
\begin{verbatim}
b_check_boolean_expression2(equivalence(LHS,RHS),_,LState,State,WF,Res,Ai,Ao) :- !,
        equiv(LR,RR,Res),
        b_check_boolean_expression(LHS,LState,State,WF,LR,Ai,Aii),
        b_check_boolean_expression(RHS,LState,State,WF,RR,Aii,Ao).
:- block equiv(-,-,-).
equiv(X,Y,Res) :-
  (  X==pred_false  -> negate(Y,Res) ;  X==pred_true    -> Res=Y
   ; Y==pred_true   -> Res=X         ;  Y==pred_false   -> negate(X,Res)
   ; Res==pred_true -> X=Y           ;  Res==pred_false -> negate(X,Y)
   ; add_error_fail(equiv,'Illegal values: ',equiv(X,Y,Res))
  ).
:- block negate(-,-).
negate(pred_true,pred_false).
negate(pred_false,pred_true).
\end{verbatim}
\end{footnotesize}
\caption{Implementation of the Equivalence Connective in \prob\label{equivalence}}
\end{figure}

%\subsection{Typical Form of the constraint}

\prob\ does not yet provide reifications for all atomic predicates.
E.g., the subset relation $\subseteq$ is not yet reified.
However, all basic predicates that appear in the Bosch application have been reified, in particular:
 \begin{itemize}
 \item $E \in P$; there are optimized reifications available for $x \in \{C_1,\ldots,C_n\}$ where $C_1,\ldots,C_n$ belong to an enumerated set or are integers,
 \item $E_1=E_2$, $E_1\neq E_2$,
 and $E_1\mathrel{\boxdot}E_2$ where $\boxdot$ is one of $<$, $\geq$, $>$, $\leq$,
 \item  Universal and existential quantifiers with small scope, which are expanded into conjunctions and disjunctions respectively.
 \end{itemize}

\begin{figure}
\begin{center}
  \includegraphics[width=8.5cm]{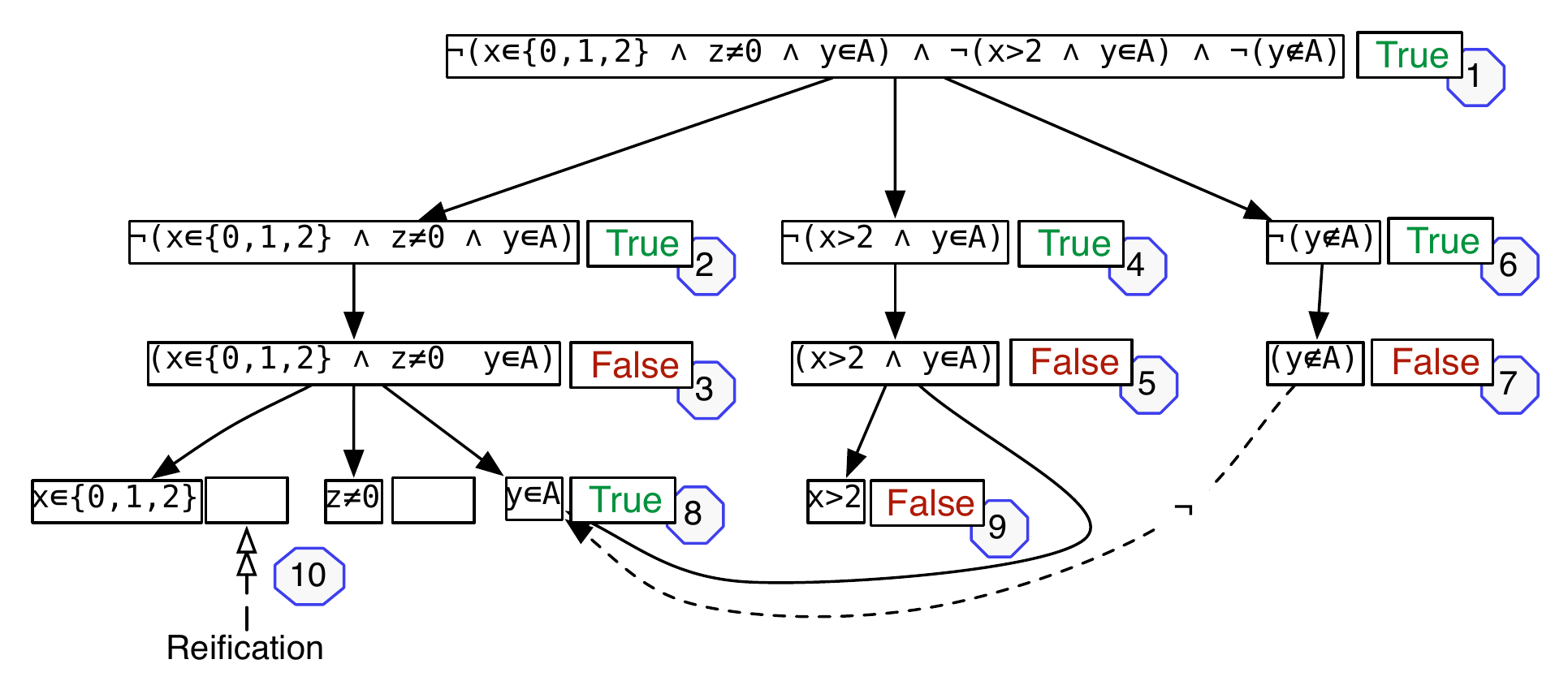} %original: {width=175pt, angle=270}
  \caption{Illustrating the \prob\ Boolean Constraint Solver and the importance of Reification (given invariant $x\geq 0$)}\label{Reification}
\end{center}
\end{figure}

\begin{example} \label{reification-example}
Figure~\ref{Reification} shows a small example, which is inspired by the deadlock constraint of the Bosch case study.
In step 1, we start by asserting that the top-level formula
 $\neg(x\in\{0,1,2\} \wedge z\neq 0 \wedge y\in A) \wedge \neg(x>2 \wedge y\in A) \wedge \neg(y\not\in A)$ is true.
We assume that an invariant $x \geq 0$ has already been asserted.
In steps 2, 4, 6, we then assert that all three sub-conjuncts must themselves be true.
In Steps 3, 5, 7 we deal with the negation.
Note that at step 7, we assert that $y\not\in A$ is false.  Due to predicate sharing,
 this immediately triggers that $y\in A$ must be true, thus forcing $x>2$ to be false in step 9.
At step 10, reification comes into play. 
Here, given the invariant $x\geq 0$, the \prob\ solver infers that $x\in \{0,1,2\}$ is true, which then forces $z \neq 0$ to be false.
In summary,
  the \prob\ kernel finds the solution $z=0$ deterministically without enumeration.
  \ignore{**
Also note that for efficiency {\em and\/} precision, the sub-formula $y\in A$ is shared.
This guarantees that if $y\in A$ is forced to be true or false (as in step 7), all other instances of $y\in A$
 are consistently updated.
Given perfect constraint propagation, the \prob\ kernel would ....
However, depending on how instantiated and complicated $A$
 and $y$ are the \prob\ kernel does not necessarily propagate this kind of information perfectly. **}
\end{example}

\label{optimizations}
 \ignore{***

\subsubsection{Optimizations} \label{optimizations}

 Sorting of predicates (cf Algorithm earlier)

 Instantiation of carrier sets and variables:
 Carrier sets: define total bijection from Cartesian product of concrete sets to abstract carrier sets.
 For instance, instantiate $S$ by two sets $S_1$ and $S_2$: $iS \in S_1\times S_2 \mathrel{\tbij} S$.
 This is used in invariants to specify instantiation of abstract variables by concrete variable $avar = iS(cvar_1 \mathrel{\mapsto} cvar_2)$;
 the same formula can be used to specifies witnesses in event refinement.
 Guards are related implicitly by means of $iS$.
 For instance, $\forall x,y\mathbin{\qdot} iS(x,y)\in A \leqv x\in\{3,5\} \lor y>8$ 
 for ``implementing'' abstract set $A$ by the condition $cvar_1\in\{3,5\} \lor cvar_2>8$. \marginpar{maybe a small example?}
 
 Symbolic treatment of (infinite) identity functions; detection of infinite closures
 
 unsat core, to detect problems (also efficiency problems);
 further application of unsat core: if no deadlock found: minimize predicate : if we have a query
 operation: everything but the query operation will be removed,...
 ***}
 
% --------------------------

\section{Case Studies}
\label{sec:casestudy}

All experiments were run on a MacBook Pro with a 3.06 GHz Core2 Duo processor, \prob\ 1.3.3 compiled
 with the 32-bit version of SICStus 4.1.3.%
%\footnote{The 64-bit version is faster; but the SICStus Tcl/Tk integration does not work yet on Mac OS.}

{\bf Standard Benchmarks}
Figure~\ref{simple} shows that the constraint-based deadlock checker (CBC) is capable of quickly
 finding deadlocks for a variety of B models (mainly taken from \cite{DBLP:conf/icfem/BendispostoL09};
 more details about the models can be found in that paper).
mondex\_m2 is the second refinement of a model of the Mondex Electronic Purse; %~\cite{Butler:2008p972};
 it contains 
5 constants, 10 variables and 6 events.
CXCC0 is a model of a congestion control protocol with 6 constants, 8 axioms,
 5 variables, 20 invariants and
 3 events.
The Siemens model %Mini Pilot was developed within the Deploy Project. It 
 is a specification of a fault-tolerant automatic train protection system, with 15 variables, 28 invariants and 10 events.
The model has an unbounded time variable and is hence infinite state.
earley\_2 is the third refinement of a model of the Earley parsing algorithm as developed by Abrial.
It contains 6 constants, 18 axioms, 5 variables, 7 invariants, and 4 complicated events. earley\_3 is
the fourth refinement of the same model.
platoon1 and platoon2 are refinement levels of a platooning system
 by Mashkoor and Jacquot. The second refinement contains 11 constants, 18 axioms, 4 variables, 6 invariants and
 7 events.
 FMCH02 is the second refinement of a file system model with 6 constants, 8 axioms,
  7 variables, 12 invariants, and 9 events. We examine the system with carrier set sizes 1 and 2.
scheduler is an Event-B translation of the scheduler from~\cite{LegeardEtAl:FME02}, %which we analyse
for a varying number of processes.
Volvo is the (B-Method) model of a vehicle function described in \cite{DBLP:journals/sttt/LeuschelB08},
 containing 15 variables and 26 events.

\begin{figure}
\begin{center}
\begin{footnotesize}
\begin{tabular}{|l||r|l||r|l|}
\cline{1-5}
Model & CBC (s) & Result & MC (s) & Result\\
\cline{1-5}
CXCC0 & 0.00 & deadlock & 0.01 & deadlock\\
earley\_2  & 0.01 & deadlock & 403.68 & no deadlock found *\\
earley\_3  & 0.01 & deadlock & 0.14 & deadlock \\
Eco\_Mch\_4 & ~~0.02 & deadlock & ~~580.28 & no deadlock found**~~~\\
FMCH02 (1) & 0.00 & deadlock & 0.01 & deadlock\\ % DEFAULT_SETSIZE 1
FMCH02 (2) & 0.53 & no deadlock~~~~~~ & 0.13 & no deadlock\\ % DEFAULT_SETSIZE 2
mondex\_m2~~~~~~~~~~ & 0.00 & deadlock & 0.01 & deadlock\\
platoon1 & 0.01 & deadlock & 0.00 & deadlock\\
platoon2 & 0.06 & deadlock & 0.03 & deadlock\\
scheduler (2) & 0.00 & no deadlock & 0.00 & no deadlock\\
scheduler (5) & 0.00 & no deadlock & 0.49 & no deadlock\\
%scheduler (7) & 0.00 & no deadlock & 1.44 & no deadlock\\
scheduler (9) & 0.00 & no deadlock & 107.18 & no deadlock***\\
Siemens & 0.00 & no deadlock & 211.28 & no deadlock**\\
Volvo & 0.20 & no deadlock & 5.16 & no deadlock\\
\cline{1-5}
\multicolumn{5}{r}{{\scriptsize *: no deadlock found after visiting 10,000 states.}}\\
\multicolumn{5}{r}{{\scriptsize **: no deadlock found after visiting 100,000 states; the system is infinite state.}}\\
\multicolumn{5}{r}{{\scriptsize ***: with hash symmetry it only takes 0.120 s to model check the system.}}
\end{tabular}
% CBC options: -p SMT TRUE
% Run using 32-bit probcli 1.3.3
% ProB Command Line Interface
%  VERSION 1.3.3-beta10 (6635:6636M)
%  $LastChangedDate: 2010-12-22 09:52:36 +0100 (Wed, 22 Dec 2010) $
%  Prolog: SICStus 4.1.3 (i386-darwin-10.4.0): Wed Sep 22 21:22:21 CEST 2010
\end{footnotesize}
% on MacBook Pro 3.06 GHz
\end{center}
\caption{Constraint-Based Deadlock Checking (CBC) vs Model Checking (MC)\label{simple}}
\end{figure}

%Mention Volvo SAE case study: all levels could be quickly proven absence of deadlocks (given settings).

{\bf Evolution: A BPEL Development} \label{BPEL-sect}
We now go into more detail for one case study, a business process for a purchase order.
%, and examine 
%how the feedback of the constraint-based deadlock checker has been used to improve the model and 
%how the performance evolves as the model evolves.
% 
The Event-B model is obtained via an automatic translation from BPEL \cite{DBLP:conf/iceccs/Ait-SadouneA09}.
Initially the model was believed to be free of deadlocks.
However, \prob\ managed to find a deadlock and it took 5 iterations to finally
 obtain a deadlock free version of the business process.
The last model has 15 events with 59 guards.
Note that the model has also driven the development of the constraint-solver; initially \prob\ was
 unable to quickly find a deadlock for the fifth version of the model.
This helped uncover an inefficiency in \prob's constraint solver.
After solving it, \prob\ now finds the deadlock almost instantaneously for the first five models
 (see Figure~\ref{bpel}).
Finally, for the sixth model, \prob\ confirms that no counter example exists for default deferred set sizes 1,2,3.

One can see in Figure~\ref{bpel} that the constraint checking time increases as the model evolves:
 as the model is improved, deadlocks get harder and harder to find.
%The same pattern appears in the Bosch application (see Section~\ref{bosch}).
(It is interesting that deadlock freedom of the fifth model was proved; however,
 it turned out that a wrong proof obligation was generated.)
In the first four models the model checker manages to find deadlocks
 also very quickly.
However, starting with the fifth version, the model checker is no longer able to provide interesting
 feedback. The out-degree for the initialisation and constant setup is just too big.

\begin{figure}
\begin{center}
\begin{footnotesize}
\begin{tabular}{|l||r|l||r|l|}
\cline{1-5}
Version & CBC (s) & Result & MC (s) & Result\\
\cline{1-5}
BPEL v1 & 0.05 & deadlock & - & no INITIALISATION~~~~~~\\
BPEL v2 & 0.05 & deadlock & 0.09 & deadlock\\
BPEL v3 & 0.04 & deadlock & 0.09 & deadlock\\
BPEL v4 & 0.03 & deadlock & 0.09 & deadlock\\
BPEL v5 & 0.13 & deadlock & 5.62 & no error found*(10)\\ %2248 states analysed
  &   & ~ & 140.82 & no error found*(100)\\ % 56200 states analysed not all transitions computed  -p MAX_OPERATIONS 100 -p MAX_INITIALISATIONS 100
  &   & ~ & 1423.56 & no error found*(1000)\\ % 562000 states analysed not all transitions computed  -p MAX_OPERATIONS 100 -p MAX_INITIALISATIONS 100
%  with max degree 100,000 no deadlock found after 97m17.719s
BPEL v6~~~~~~~ & ~~0.37 & no deadlock~~~~~ & 5.61 & no error found*(10)\\ % 2248 states analysed
  &   & ~ & ~~~~~140.89 & no error found*(100)\\ % 56200 states analysed not all transitions computed  -p MAX_OPERATIONS 100 -p MAX_INITIALISATIONS 100\\
\cline{1-5}
\multicolumn{5}{r}{{\scriptsize *(X): not all transitions computed; maximum out-degree X}}
\end{tabular}
% CBC options: -p SMT TRUE
% Run using 32-bit probcli 1.3.3
% ProB Command Line Interface
%  VERSION 1.3.3-beta10 (6635:6636M)
%  $LastChangedDate: 2010-12-22 09:52:36 +0100 (Wed, 22 Dec 2010) $
%  Prolog: SICStus 4.1.3 (i386-darwin-10.4.0): Wed Sep 22 21:22:21 CEST 2010
% on MacBook Pro 3.06 GHz
\end{footnotesize}
\end{center}
\caption{Comparing Constraint-Based Deadlock Checking (CBC) and Model Checking (MC) on multiple versions of the same model\label{bpel}}
\end{figure}

{\bf The Bosch Cruise Control Application} \label{bosch}
The main motivation for this work was the deadlock checking for a cruise control system modelled by
 Bosch within the Deploy project.
Indeed, proving absence of deadlocks is crucial in this case study \cite{DeployD19}, as it means that the engineers have thought of every possible scenario.
In other words, a deadlock means that the system can be in a state for which no action was foreseen by the engineers.

The model contains many levels of refinement and the
 particular machine of interest %(CrCtrl\_Comb2Final)
 is very big: it contains 78 constants with 121 axioms, 62 variables with 59 invariants
  and has 80 events with 855 guards (39 of them are disjunctions, containing 17 more conjuncts nested
   inside).
Of the 140 variables and constants,
  4 have $2^{13}=8{,}192$ possible values, 11 have $2^{32}$ %=$4,294,967,296$
   possible values, one has $2^{52}$, %= $4{,}503{,}599{,}627{,}370{,}496$, 
   another one has
  $2^{65}$, %$=36{,}893{,}488{,}147{,}419{,}103{,}232$
   and 79 variables or constants have infinitely many possible values
  (or so many that they cannot be represented as a floating number).
 The resulting deadlock-freedom proof obligation is very big: when printed it takes 34 pages of A4 using 9-point Courier.
Initially, the Rodin toolset also had trouble loading this proof obligation resulting in a ``Java Heap Space Error''.
Furthermore, even after successfully loading the proof obligation into the Rodin proving environment,
  it is very tedious for a user to try discharging the proof obligation and the information obtained from the failed proof attempt is not very useful. 
 
Here \prob's constraint-checking feedback has been very valuable: it provides the Bosch engineers with a concrete scenario which has not yet
 been anticipated and allows them to modify the model accordingly.
\prob\ can then be run again on the modified model, until no more deadlock can be found.
One can then switch to the Rodin provers to discharge the proof obligation.
(For a smaller version of the model this was actually very successful: the newPP prover was then able to automatically
  discharge the proof obligation).

%Adaptive Cruise Control description
%\rem{http://www.kfztech.de/kfztechnik/sicherheit/acc.htm ; there's also the Kraftfahrtechnisches Taschenbuch by Bosch}

%\rem{We could present an anonymised version of the model, just explaining the typical predicates and number of events.}

%Figure~\ref{bosch_exp} contains a summary of the experimental results for
The latest version of \prob\ takes from 1.07 to 2.32 seconds for finding deadlocks
 for various versions of the Bosch model for a particular
  predicate of interest ({\tt\small Counter=10}).
\ignore{***
Indeed, the model is conjoined with a simple controller (which is not encoded in Event-B), meaning that
 not all sequences of operations are actually feasible.
In this application, the engineers were only interested in deadlocks that could appear when the Counter variable
 was set to 10.
***} 
Also note that loading and type checking the model takes a considerable amount of time.
For example, the Prolog representation of the abstract syntax tree takes about 7.5 MB on disk.
The total time for finding a deadlock, including loading, type-checking, building the constraint
  and constraint solving,
  hence takes from 9.98 to 11.92 seconds.

Note that model checking of these models was not really successful.
E.g., for the latest version %v9
 of the model the model checker requires 50.41 seconds in total %36.61 seconds mc time (total time 50.41 seconds)
 to find a deadlock.
Unfortunately, this is not a deadlock that is of interest to the Bosch engineers (we have {\tt\small Counter=1} for the
 deadlock state).
When searching specifically for deadlocks with {\tt\small Counter=10}, the model checker failed to find
 a counter example after running for
 %14 minutes and 26.90 seconds total runtime, % (845.20 seconds model checking time),
 % using the default settings for the maximum out-degree and checking in total 100 nodes.
%With an out-degree of 20, the model checker ran for
 almost 4 hours (with a maximum out-degree of 20).

In summary, the result of this case study has been very encouraging.
We have managed to solve big deadlock constraints %, spanning 34 pages of A4,
 of a real industrial application.
The obtained deadlock counter examples have been extremely useful to the engineers,
 helping them to improve the model.

\section{Related Work and Conclusion}

As far as constraint solving for sets is concerned we would like to mention setlog
 \cite{DovierEtAl:Toplas00}, which is unfortunately no longer maintained.
Setlog has certain restrictions (e.g., interval bounds must be known values, $x\in y..6$ is not accepted)
 and does not seem to cater for reification, which we used for effective integration into a boolean constraint solver.
We conducted one comparison, solving the N-Queens problem for $n=14$: setlog 4.6.14 took about 40 seconds to find the
 first solution, compared with 0.03 seconds for \prob.
Another tool of interest is BZ-TT \cite{LegeardEtAl:FME02}. %\cite{Ambert:FATES02}. % (building on CLPS-B \cite{Bouquet:TACAS02}).
This tool can also be used for constraint solving, and has been used for test-case generation,
 but its support for B quite limited (e.g., it does not support set comprehensions
 % nor lambda abstractions 
  nor refinement).
Also, we were unable to load and solve the BPEL deadlock constraints (Sect.~\ref{BPEL-sect}) 
 nor solve an N-Queens puzzle with BZ-TT.
% Example from Figure~\ref{Reification} can be solved, but more complicated formulas cannot.
Two more animation tools for B are AnimB and Brama.
As we have shown in \cite{LeuschelEtAl:STS09}, none of them are capable
 of dealing with more sophisticated constraints.
The same is true of the TLC model checker \cite{DBLP:conf/charme/YuML99} for TLA$^{+}$.
Alloy \cite{Jackson:Alloy} on the other hand can be used for constraint solving and has been used in at least one instance
 for deadlock checking \cite{Dillon06developingan}.

Our deadlock constraint (DLN) is often already very close to being in conjunctive
 normal form (CNF).
As such, one may wonder whether SAT or SMT technology could have been employed for our application.

{\bf SAT}
In \cite{DBLP:conf/flops/HoweK10} Howe and King present a Prolog SAT solver which uses co-routines to
 implement unit propagation efficiently and elegantly.
The \prob\ boolean constraint solver also achieves unit-propagation, but is not optimized for CNF.
In particular, \prob\ creates a variable for {\em every\/}
  subformula and attaches co-routines to it, whereas \cite{DBLP:conf/flops/HoweK10}
  uses a clever scheme tailored for CNF to wait only on two variables per clause.
Still, \prob\ can solve some non-trivial SAT problems when encoded in B.
E.g., for the most complicated SATLIB example in \cite{DBLP:conf/flops/HoweK10}
 (flat200-90 with 600 Boolean variables and 2237 Clauses)
 \prob\ takes 3.27 seconds % latest ProB 1.3.4-beta1 takes 1.86 seconds
  to find the first solution (successive solutions are then found very quickly).
The Prolog SAT solver from \cite{DBLP:conf/flops/HoweK10} takes only 0.13 seconds to solve this
example and minisat \cite{DBLP:conf/sat/EenS03}
is even faster (about 0.01 seconds).%
\footnote{Surprisingly, the CLP(B) solver from SICStus Prolog 4 using BDDs runs out of memory after about 5 minutes. 
The very latest beta version 1.3.4 of \prob\ solves it in 1.85 seconds.}
%Note that on the SAT problem generated by Kodkod, minisat takes 1.50 seconds.
Still, \prob\ is working directly on a high-level formalism: for the usual applications of \prob, the number
of clauses is much smaller than for SAT encodings.
%\footnote{As we have seen in Section~\ref{bosch}, \prob\ is capabable of dealing with 30 page
% high-level formulas.}
\prob\ also has to deal with issues such as potentially undefined expressions, which leads
 to performance penalties and makes a CNF encoding less appealing.
%In particular for issues concerning well-definedness, we currently want to avoid expanding
% the B formulas into CNF.
Granted, better encodings are available to solve pure SAT problems, but in the setting of B and Z
 it is unclear whether an approach such as \cite{DBLP:conf/flops/HoweK10} would pay off.

{\bf SMT}
Compared to SMT solving our constraint-solving approach uses static ordering and
 is capable of theory propagation via reification
 (see Example~\ref{reification-example}), but is lacking one important feature:
 clause learning.
 % Partition example, SMT Learning vs deterministic constraint solving, ...
However, to apply SMT solvers to our deadlock formulas we need support for set theory, relations and functions.
Such support is not yet openly available.
The company Systerel is currently developing a translator from B to SMTLIB based on \cite{DBLP:conf/asm/Deharbe10}.
We have used a beta version of the translator on the BPEL examples from Sect~\ref{BPEL-sect}.
Unfortunately, we were not able to use the VeriT SMT solver due to a bug in the translator. %, related to existential quantification.
 %on fourth BPEL model and simple and full version of DLF:
 % "veriT could not parse rodin\_sequent.smt".
 %This also happens if we use veriT as the preprocessor.
 %Problem proably due to existential quantifier (note: we applied veriT after applying our existential
 % quantifier simplification and after Rodin simplification).
We were able to run CVC3-2.2 on the deadlock constraint of v4: it  ran for 30 seconds without displaying a result, after which the
Rodin time-out aborted the process.
Note that \prob\ takes 0.03 seconds to find a counter example.
So far we have also not been able to use Kodkod \cite{DBLP:conf/tacas/TorlakJ07} high-level interface to SAT (used by Alloy)
 to solve the deadlock constraints for this example.
%
%By and large, currently we have not been able to apply SAT or SMT technology to solve the
% deadlock constraints in our applications.
Nonetheless, we are still investigating this research avenue further. % and are developing a translator from B to Kodkod.

In conclusion, we have presented a way to validate deadlock freedom of B and Event-B models using a constraint-solving approach.
We have shown an algorithm for constraint-based deadlock checking and believe further significant will be possible by combining
 constraint-solving with theorem proving. %\marginpar{OK??} Ja
We have compared the approach with model checking.
The implementation of the \prob\ constraint solver has been presented and its performance has been
 evaluated on a series
of benchmarks and one industrial application.
Summing up, the \prob\ constraint solver written in Prolog manages to solve very large deadlock constraints
 in practical examples.
The feedback obtained by our new technique has been very useful to engineers.
Thus far, we have been unable to apply SAT, SMT or model checking technology on the industrial application.

% --------------------------
 
\subsection*{Acknowledgements}
{\small 
We are grateful for the fruitful interactions with Rainer Gmehlich, Katrin Grau and Felix L\"osch from Bosch.
Thanks to David Deharbe, Yoann Guyot and Laurent Voisin for giving us access to the
 B to SMT plugin.
We thank Yamine A\"{\i}t Ameur and Idir A\"{\i}t-Sadoune for the BPEL case study and the
 prolific interactions. Thanks to Daniel Plagge for implementing the record detection.
 % and improving the type checking performance.
Finally, part of this research has been funded %research
by the EU FP7 project 214158: DEPLOY.
% (Industrial deployment of advanced system engineering methods for high productivity and dependability).
}

% --------------------------

\bibliographystyle{acmtrans}

\bibliography{michael,stefan}

\ignore{***
%\newpage
\appendix

\section{Deadlock Counter Example (For Referees)}

%Figure~\ref{Bosch-Counter}
Below is a graphical visualization of the core of the deadlock constraint
 for the ninth version of the Bosch model.
It only shows the constraints related to the guards of the events and their truth values in
 the counter example found by \prob.

%\begin{figure}
\begin{center}
  \includegraphics[width=1.5cm]{Figures/Counter1} 
  \includegraphics[width=1.5cm]{Figures/Counter2} 
  \includegraphics[width=1.5cm]{Figures/Counter3} 
  \includegraphics[width=1.5cm]{Figures/Counter4} 
%  \caption{Illustrating the Counter Example Found by \prob\ for Bosch v9}\label{Bosch-Counter}
\end{center}
%\end{figure}
**}

\end{document}